\documentclass{article}

 \usepackage[main, final]{neurips_2025}

\usepackage[utf8]{inputenc} % allow utf-8 input
\usepackage[T1]{fontenc}    % use 8-bit T1 fonts
\usepackage{hyperref}       % hyperlinks
\usepackage[most]{tcolorbox}
\usepackage{url}            % simple URL typesetting
\usepackage{booktabs}       % professional-quality tables
\usepackage{amsfonts}       % blackboard math symbols
\usepackage{nicefrac}       % compact symbols for 1/2, etc.
\usepackage{microtype}      % microtypography
\usepackage{graphicx}
\usepackage{xcolor}         % colors
\usepackage{enumitem}
\title{AgenticRS-Architecture: System Design for Agentic Recommender Systems}

\author{%
\textbf{Hao Zhang}$^{1*}$,
\textbf{Jinxin Hu}$^{1*}$,
\textbf{Hao Deng}$^{1*}$,
\textbf{Lingyu Mu}$^{2}$,
\textbf{Shizhun Wang}$^{1}$,
\textbf{Yu Zhang}$^{1}$,
\textbf{Xiaoyi Zeng}$^{1}$\\[4pt]
$^{1}$Alibaba International Digital Commerce Group, Beijing, China\\
$^{2}$University of Chinese Academy, Beijing, China\\[4pt]
\texttt{jinxin.hjx@alibaba-inc.com},
\texttt{denghao.deng@alibaba-inc.com},
\texttt{mulingyu@iie.ac.cn},\\
\texttt{zh138764@alibaba-inc.com},
\texttt{shaoan.wsz@taobao.com},
\texttt{daoji@alibaba-inc.com},\\
\texttt{yuanhan@taobao.com}%
\thanks{Equal contribution.}%
}

\begin{document}

\maketitle

\begin{abstract}
  This paper presents AutoModel, an agentic architecture for the full lifecycle of industrial recommender systems. Instead of a fixed recall–ranking pipeline, AutoModel organizes recommendation as a set of interacting evolution agents with long term memory and self improvement capability. We instantiate three core agents along the axes of models, features, and resources: AutoTrain for model design and training, AutoFeature for data analysis and feature evolution, and AutoPerf for performance, deployment, and online experimentation. A shared coordination and knowledge layer connects these agents and records decisions, configurations, and outcomes. Through a case study of \textit{paper\_auto\_train}, we show how AutoTrain automates paper driven model reproduction by closing the loop from method parsing to code generation, large scale training, and offline comparison, reducing manual effort for method transfer. We argue that AutoModel enables locally automated yet globally aligned evolution of large scale recommenders and can be generalized to other AI systems such as search and advertising.
\end{abstract}

\section{Introduction}
Recommender systems are core infrastructure for content feeds, short video, and e commerce \cite{wang2021survey, wang2024rethinking, lin2024enhancing,wang2025home,mu2025trust,deng2025csmf}. Industrial systems have evolved from collaborative filtering\cite{resnick1994grouplens} to deep models and large pretrained models \cite{mu2026masked,cobra,xing2025reg4rec,deng2025csmf,guo2017deepfm,zhou2025onerec,lian2018xdeepfm}, and from single models to multi stage pipelines with recall, coarse ranking, fine ranking, and reranking \cite{zhou2018din,mu2025trust}. Despite these advances \cite{wang2021survey, wang2024rethinking, lin2024enhancing}, both pipelines and recent end to end architectures remain largely static: modules are fixed at design time, treated as black boxes, and improved mainly through manual hypothesis, model changes, and retraining \cite{gray1984vector,rpg,zhou2025onerec, mu2025synergistic,wang2024learnable,lee2022autoregressive,wang2023generative}. This human centric, structurally rigid paradigm is increasingly a bottleneck under heterogeneous data and rapidly changing business goals.

Building on earlier work on Agentic Recommender Systems (ARS) \cite{yao2022react,202603.2003}, which modeled closed loop, independently evaluable, and evolvable components as recommender agents, this paper focuses on system architecture. We propose AutoModel, an agentic framework for the full lifecycle of recommender models. AutoModel treats data analysis and feature engineering, model design, training and offline evaluation, deployment and inference optimization, and online experimentation as one coherent process, and organizes it through three core agents coordinated by a shared orchestration and knowledge layer. AutoFeature handles data profiling, feature candidate generation and selection, and feature pipeline management. AutoTrain handles method extraction from papers and requirements, code modification or synthesis, training job execution, and offline result analysis. AutoPerf handles training and inference performance, deployment and rollback, and A/B experiment and monitoring.

AutoModel is designed to align agent boundaries with lifecycle stages, to expose explicit shared state and task orchestration across agents and business systems, and to embed evolution mechanisms with inner and outer rewards at the architectural level. 
Our contributions are be sumarized as follows:
\begin{itemize}[ topsep=0pt, leftmargin=*]
    \item We introduce AutoModel as an end to end agentic architecture for recommender model lifecycle management. 
    \item We define the structure and responsibilities of AutoFeature, AutoTrain, and AutoPerf under the ARS principles of closed loop functionality, independent evaluation, and evolvable decision space. 
    \item We provide a case study of AutoTrain on paper driven model reproduction and refinement, showing how a self improving model agent can be instantiated in practice.
\end{itemize}

\section{Model Iteration Analysis}
This section starts from the full production and iteration lifecycle of industrial recommender models, analyzes the key stages and pain points, and derives the requirements that AutoModel must satisfy. We treat research and engineering activities as one coherent system so that later agent design and interfaces are grounded in real workflows.

\subsection{Model lifecycle and structural issues}
In production, recommender models evolve through a long cross team pipeline. Practitioners diagnose issues from logs and metrics, search the literature and design variants, adjust features and data construction, implement and tune training jobs, then compress and deploy models for online A/B tests before deciding rollout or rollback. This cycle is slow and brittle. It is dominated by manual decisions, scattered across reports, documents, scripts, and dashboards, so past successes and failures are hard to reuse. Objectives differ by stage, causing local optimizations that may not improve overall business metrics. Automation tools such as hyperparameter search or LLM based code generation remain isolated within individual platforms and do not close the loop.

From an agentic view, two gaps are central. Lifecycle state and knowledge lack a unified representation, and crucial choices such as feature selection and model configuration are not treated as independently evaluable decision units. AutoModel addresses these gaps by reifying such decisions as internal agents that share a common architecture and can coevolve over time.

\subsection{Core requirements for the AutoModel architecture}
The above analysis yields several requirements. The system should maintain a unified representation of lifecycle state and knowledge, where problem statements, model variants, feature and data settings, training runs, and offline or online results are stored as queryable, linked metadata rather than scattered documents and logs. The architecture should assign agent oriented responsibilities to key decision stages, with dedicated agents for data and features, model design and training, and performance and deployment. AutoFeature, AutoTrain, and AutoPerf instantiate these roles and each builds its own perception–decision–execution–feedback loop behind standard interfaces.

The architecture also needs an explicit decision space and stable interfaces for evolution methods. For model related agents, it should specify which structures and configurations are open to change, so that reinforcement learning or search can tune discrete configurations and large language models can propose new architectures and training schemes. These updates must feed into a common train–evaluate–select loop instead of remaining isolated scripts.

\section{AutoModel Architecture}

\subsection{Overview}
\begin{figure}[htbp]

  \includegraphics[width=1.0\textwidth]{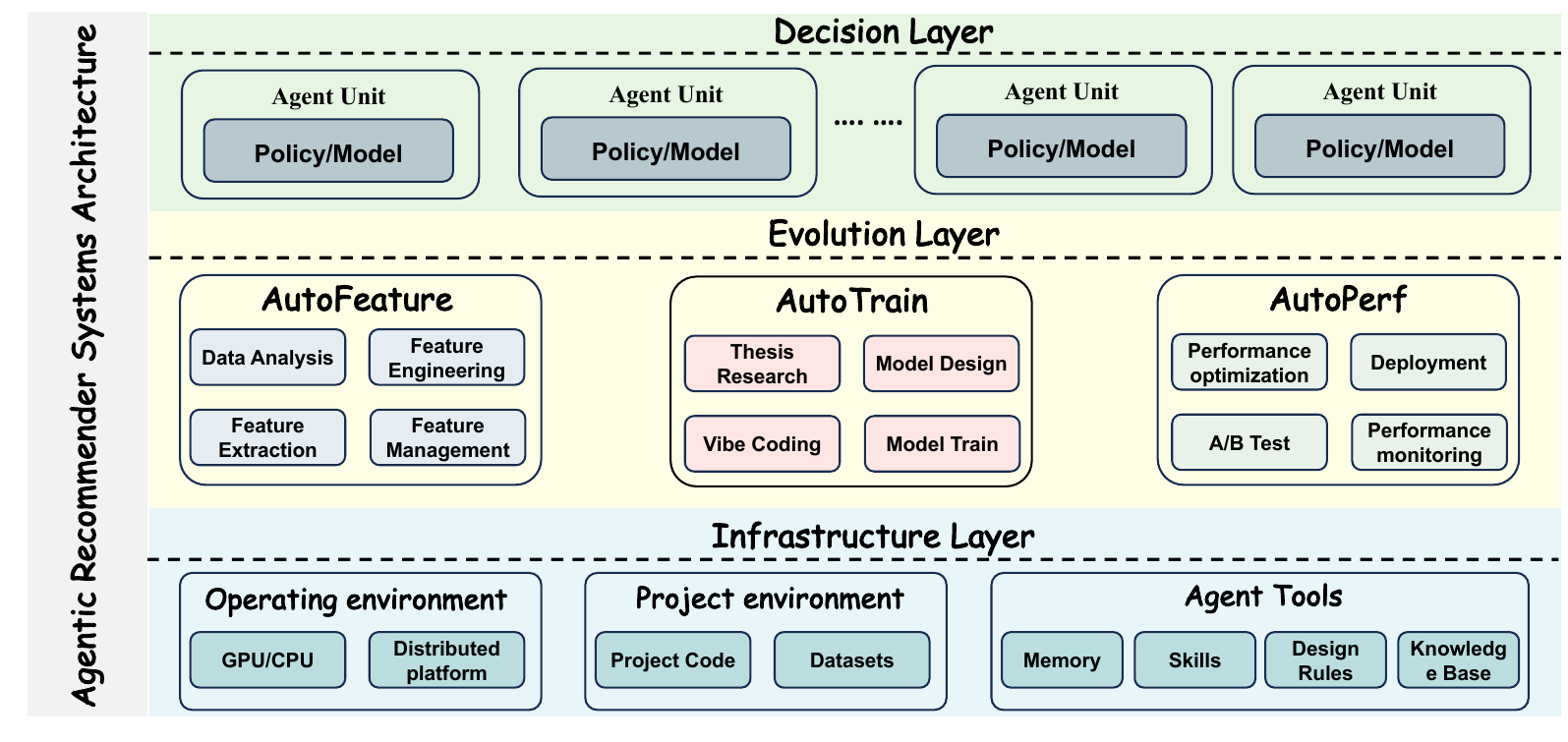}
  \vspace{-5pt}
  \caption{
  The architecure of agentic recommender systems.
  }
  % \Description{xx.}
  \label{fig1}
  \vspace{-0.5cm} %调整图片与上文的垂直距离
\end{figure}

AutoModel replaces the traditional split between recall, ranking, policy modules and tooling platforms with a set of agents that have clear responsibilities, independent evaluation, and self evolution capability. The overall system can be viewed as a multi layer agent graph as shown in Figure \ref{fig1}. At the \textbf{decision layer}, online recommendation agents respond to user requests and play roles similar to recall, ranking, reranking, and policy control, but their internal structure and composition are allowed to evolve over time. At the \textbf{evolution layer}, agents for data analysis, model design, training, and deployment continuously generate and update decision agents based on logs and rewards. At the \textbf{infrastructure layer layer}, infrastructure for task orchestration and knowledge storage maintains shared state and experience. 

\subsection{Core Evolution Agents}
AutoModel centers its evolution on three loops for data and features, model design and training, and performance and deployment, realized by AutoFeature, AutoTrain, and AutoPerf, which jointly drive the long term evolution of decision agents.

\textbf{AutoFeature} manages data usage and feature representation. It connects to logs and content, diagnoses data quality and target relevant slices, proposes feature construction and selection schemes, and registers validated features as shared representations. Unlike one off scripts, it acts as a persistent agent that updates feature sets based on online feedback, thus changing what downstream models can observe.

\textbf{AutoTrain} evolves model architectures and training procedures. It ingests problem descriptions, feature plans, and external methods, produces model and training configurations, edits or synthesizes code, submits large scale training jobs, and analyzes offline results. Recall and ranking models are treated as decision agent instances managed by AutoTrain, which can modify, replace, or recombine them according to feedback.

\textbf{AutoPerf} brings candidate decision agents online and optimizes behavior at inference and experimentation time. It allocates training and serving resources, configures parallelism and compression, and manages deployment, staged rollout, and A/B experiments, feeding business and risk signals back to AutoTrain and AutoFeature. In this way, how to deploy and how to experiment become explicit optimization targets.

\subsection{Coordination and Knowledge Layers}
To ensure that evolution agents and decision agents form a coherent system rather than isolated tools, AutoModel introduces a coordination layer and a knowledge layer. The coordination layer decomposes long term goals or concrete tasks into cross agent workflows. Unlike fixed pipelines, these workflows can adapt to system state and historical experience, sometimes emphasizing AutoTrain for major architectural changes, and other times relying more on AutoFeature or AutoPerf for feature or traffic adjustments. It maintains explicit task graphs and state machines that govern agent invocation order, failure recovery, and termination, making system behavior interpretable to both humans and agents. The knowledge layer provides shared memory for all agents. It stores problem definitions, data analysis results, model and feature configurations, training and evaluation logs, and online experiment conclusions. New variants and experimental outcomes are written to this layer in each iteration. Subsequent decisions by AutoFeature, AutoTrain, and AutoPerf read from this shared knowledge to avoid repeating unproductive directions and to expose cross task patterns. In line with the inner and outer reward view, the knowledge layer also carries reward signals and credit assignment records across agents.

\section{Evolution Agents}
This section introduces the three core evolution agents in AutoModel: AutoTrain, AutoFeature, and AutoPerf. They are long lived decision units rather than one off tools and drive the self evolution of the recommender system along model, representation, and system performance dimensions.

\subsection{AutoTrain: Evolution in the Model Space}
AutoTrain decides which model architectures and training procedures to use under given business goals and data conditions. It maintains a family of model variants, proposes new candidates from performance signals, historical models, feature plans, and methods from papers, maps them to the existing codebase, and runs a standardized loop of code modification, training, and offline evaluation. Over time it accumulates configuration–performance experience to narrow search. AutoTrain outputs evaluated model variants and does not directly control deployment.

\subsection{AutoFeature: Evolution in the Representation Space}
AutoFeature determines what information models observe. It continuously analyzes logs and data warehouses to monitor distributions and shifts, and, given current goals and feedback from AutoTrain and AutoPerf, proposes new features or encodings and retires costly or ineffective ones. It follows a loop of diagnosing data issues, generating feature candidates using knowledge of past successes and failures, and relying on AutoTrain to assess marginal gains. Its outputs are feature plans and evaluations, turning representation design into a continual decision process.

\subsection{AutoPerf: Evolution in the Resource and Risk Space}
AutoPerf decides how to train and deploy candidate models under resource and risk constraints. It aggregates business requirements such as compute budget, latency, availability, and risk tolerance with internal signals such as model complexity, convergence, feature cost, and prior experiment outcomes. Based on these, it allocates training and serving resources, selects parallelism and compression strategies, and configures A/B experiments including traffic split and rollback policies. Its decisions are evaluated by online metrics and risk events, and AutoPerf enforces risk boundaries while balancing exploration and exploitation for the whole system.

\section{Case Study: paper\_auto\_train}
\begin{figure}[t]

  \includegraphics[width=1.0\textwidth]{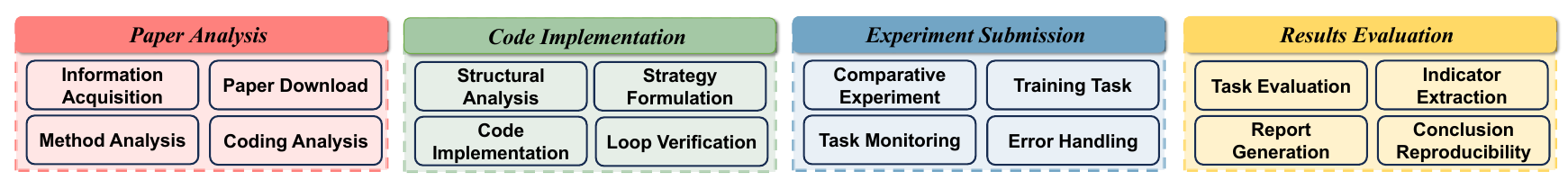}
  \vspace{-5pt}
  \caption{
  The pipeline of \textit{paper\_auto\_train}.
  }
  % \Description{xx.}
  \label{fig2}
  \vspace{-0.5cm} %调整图片与上文的垂直距离
\end{figure}

This section instantiates AutoTrain through an internal capability called \textit{paper\_auto\_train} as shown in \ref{fig2}, which automates a frequent yet costly task in industrial recommender systems: reproducing research papers and adapting their methods for model iteration. 

\subsection{Background and Objective}
In production recommenders, reproducing published methods on internal data is a major source of model evolution, but the traditional workflow is manual and slow, from reading papers and understanding methods to locating or writing code, adapting data, running training, and comparing results. Under the AutoModel perspective, we reframe this as a standard AutoTrain subtask. Given a paper or method description, AutoTrain should automatically perform method parsing, code adaptation, training, and result comparison, and store the process and conclusions in the knowledge layer as reusable assets. The paper\_auto\_train pipeline is a concrete realization of this goal.

\subsection{Pipeline Overview}

\subsubsection{Phase 1: Paper Parsing and Method Abstraction}
The pipeline starts from a user provided paper hint such as a title, identifier, HTML link, or PDF URL. AutoTrain uses a paper parsing sub agent backed by an LLM to fetch the content when needed and extract a structured method description that covers the target problem, key modeling ideas, high level architecture, input and output requirements, loss functions, training strategies, and reported results. These fields are written into the knowledge layer and aligned with current business issues and baseline models, providing a semantic blueprint that guides later code changes and evaluation.

\subsubsection{Phase 2: Code Analysis and Model Implementation}
Given the method blueprint, AutoTrain maps it onto the existing system. A code analysis and rewriting sub agent inspects the target repository, identifies relevant components such as data input, model definitions, training loops, and evaluation scripts, and decides where to insert or replace modules according to the Phase 1 description. The LLM translates symbols and pseudocode from the paper into concrete APIs and modules, generating initial edits that are checked by static analysis and basic tests. AutoTrain maintains both a baseline script and a paper based variant that share the same data and feature pipelines, ensuring controlled comparison.

\subsubsection{Phase 3: Training Submission and Monitoring}
With runnable code, AutoTrain submits training jobs through the training platform and monitors their lifecycle instead of simply launching scripts once. It tracks job identifiers and configurations, periodically inspects logs for signals such as loss curves, gradient stability, and resource usage, and when anomalies are detected it generates explanations and remediation suggestions such as adjusting batch size or learning rate. For issues, AutoTrain updates configurations and resubmits jobs automatically; for more complex cases it escalates diagnostics to human engineers. When training finishes, AutoTrain collects metadata on duration, resource consumption, and model size and stores them with offline metrics in the knowledge layer for future configuration and cost modeling.

\subsubsection{Phase 4: Result Comparison and Reporting}
After training both baseline and paper variants, AutoTrain invokes a unified evaluation pipeline on the same validation and replay data to compute offline metrics such as AUC, NDCG, recall, and scenario specific slices. It then performs a structured comparison to identify where the paper method improves or degrades performance across user groups, content types, and business scenarios. The findings are compiled into a report with tables, plots, textual summaries, and recommendations on whether further engineering or feature or loss adjustments are warranted. Methods that show no practical value are recorded as negative cases with reasons in the knowledge layer to avoid redundant future attempts. In this way, paper\_auto\_train turns paper reproduction from a manual, experience driven procedure into a well scoped, repeatable, and self improving AutoTrain loop.

\section{Experiments}

In this section, we present a concrete execution trace of the \texttt{paper\_auto\_train} agent on a real task: reproducing the NeurIPS 2025 Best Paper \emph{Gated Attention for Large Language Models: Non-linearity, Sparsity, and Attention-Sink-Free} \cite{qiu2025gated}. The goal is to automatically parse the method, implement the key architectural changes, submit baseline and experimental training jobs, and prepare evaluation for offline comparison.

\subsection{Step 1: Discovering and Summarizing the Paper}

Given a user request ``reproduce the Gated Attention paper'', the agent first retrieves basic metadata and summarizes the core idea.

\begin{tcolorbox}[
  colback=gray!15,
  colframe=black!30,
  boxrule=0.4pt,
  left=4pt,right=4pt,
  top=3pt,bottom=3pt,
  enhanced,
  sharp corners,
  before skip=3pt,
  after skip=3pt
]
\small
Agent: Let me help you reproduce this paper. I will first read the paper to understand its core method and technical details.\\
Agent: The PDF link is not directly accessible. I will try the OpenReview web page and then fall back to web search if needed.\\
Agent: I have found the paper: \emph{Gated Attention for Large Language Models: Non-linearity, Sparsity, and Attention-Sink-Free}, a NeurIPS 2025 Best Paper by the Qwen Research Team.\\
Agent: The key idea is to add a head-specific, query-dependent sigmoid gating mechanism (G1) after Scaled Dot-Product Attention (SDPA) to remove attention sink and improve training stability and scalability.
\end{tcolorbox}

Based on this, \texttt{paper\_auto\_train} stores a structured summary including title, venue, and a list of key contributions (G1 gating, mitigation of attention sink, improved stability, and activation sparsity) into the shared knowledge layer.

\subsection{Step 2: Planning and Implementing Code Changes}

Next, the agent analyzes the existing recommendation codebase, locates the attention module, and plans the minimal set of changes needed to instantiate the Gated Attention variant alongside the baseline.

\begin{tcolorbox}[
  colback=gray!15,
  colframe=black!30,
  boxrule=0.4pt,
  left=4pt,right=4pt,
  top=3pt,bottom=3pt,
  enhanced,
  sharp corners,
  before skip=3pt,
  after skip=3pt
]
\small
Agent: I will analyze the current project code structure and identify the model entry points and configuration files.\\
Agent: According to the paper, I need to add a head-specific, query-dependent sigmoid gate after SDPA. I will create a new experimental model variant.\\
Agent: Creating the experimental model and training configuration, as well as submission and evaluation scripts for the Gated Attention version.\\
Agent: I will run a syntax check before submitting any training jobs.
\end{tcolorbox}

The resulting implementation includes new model and configuration files and a small change to the multi-head attention module:

\begin{tcolorbox}[
  colback=gray!15,
  colframe=black!30,
  boxrule=0.4pt,
  left=4pt,right=4pt,
  top=3pt,bottom=3pt,
  enhanced,
  sharp corners,
  before skip=3pt,
  after skip=3pt
]
\small
\verb|attn_output = F.scaled_dot_product_attention(q, k, v, attn_mask=attn_mask)|\\
\verb|gate_values = self.gate_proj(q)        # [batch, nhead, seq_len, 1]|\\
\verb|gate_values = torch.sigmoid(gate_values)|\\
\verb|attn_output = attn_output * gate_values|
\end{tcolorbox}

\subsection{Step 3: Submitting and Tracking Training Jobs}

After the code is ready, the agent submits both baseline and Gated Attention variants to the training platform and keeps a concise record that is later used by AutoPerf and by human operators.

\begin{tcolorbox}[
  colback=gray!15,
  colframe=black!30,
  boxrule=0.4pt,
  left=4pt,right=4pt,
  top=3pt,bottom=3pt,
  enhanced,
  sharp corners,
  before skip=3pt,
  after skip=3pt
]
\small
Agent: Syntax check passed. I will submit the training jobs: first the Base version, then the Gated Attention experimental version.\\[2pt]
\textbf{Training job submission records}\\
\begin{tabular}{lll}
\textbf{Version} & \textbf{Task name}                 & \textbf{Task ID} \\
Base             & jiehong\_recis\_pg\_pool\_v2       & 3872848b...0a1e16 \\
Gated Attention  & jiehong\_recis\_pg\_pool\_v2\_gated & 0844ee4a...91aa20e \\
\end{tabular}\\[2pt]
Agent: I will monitor these jobs and, once they finish, trigger the evaluation scripts to compare the baseline and Gated Attention models.
\end{tcolorbox}

This experiment illustrates how \texttt{paper\_auto\_train} decomposes a high level request (reproduce a paper) into a sequence of controllable steps, and how short, human readable execution traces can be logged at each step to make the agent workflow transparent and auditable.

\section{Conclusion}

Building on the ARS design principles, this work proposes AutoModel, a concrete architecture for next generation recommender systems that treats the system as a set of interacting evolution agents rather than a fixed recall–ranking pipeline. Along three dimensions of models, features, and resources, we design three core agents, AutoTrain, AutoFeature, and AutoPerf, and organize them through a shared coordination and knowledge layer into a system with long term memory and self evolution capability. Using paper\_auto\_train as a case, we show how AutoTrain turns paper parsing, code adaptation, training, and evaluation into a repeatable closed loop, significantly reducing manual effort for method transfer and model iteration. With layered rewards and multi agent coevolution, AutoModel offers a practical path toward locally automated yet globally aligned industrial recommenders, and can be extended to search, advertising, and other complex AI systems.

\bibliographystyle{plain}
\bibliography{references}

\end{document}